\documentclass[pdflatex,sn-mathphys-num]{sn-jnl}% Math and Physical Sciences Numbered Reference Style
\usepackage{graphicx}%
\usepackage{multirow}%
\usepackage{amsmath,amssymb,amsfonts}%
\usepackage{amsthm}%
\usepackage{mathrsfs}%
\usepackage[title]{appendix}%
\usepackage{xcolor}%
\usepackage{textcomp}%
\usepackage{manyfoot}%
\usepackage{booktabs}%
\usepackage{algorithm}%
\usepackage{algorithmicx}%
\usepackage{algpseudocode}%
\usepackage{listings}%
\usepackage{float, subfigure}
\usepackage{epstopdf}
\usepackage{hyperref}
\usepackage{xcolor,enumitem,rotating}

\theoremstyle{thmstyleone}%
\theoremstyle{thmstyletwo}%

\theoremstyle{thmstylethree}%

\newcommand{\cynthia}[1]{\textcolor{black}{#1}}

\begin{document}

\title[Article Title]{UKAN-EP: Enhancing U-KAN with Efficient Attention and Pyramid Aggregation for 3D Multi-Modal MRI Brain Tumor Segmentation}


\author{\fnm{Yanbing} \sur{Chen}$^\dag$}

\author{\fnm{Tianze} \sur{Tang}$^\dag$}

\author{\fnm{Taehyo} \sur{Kim}}

\author{\fnm{Hai} \sur{Shu}$^*$}

\affil{Department of Biostatistics, School of Global Public Health, 
New York University, New York, NY, USA}

\affil{$\dag$ Equal contributions}
\affil{$*$ Corresponding author. Email: \url{hs120@nyu.edu}}

\affil{\\ $~~~$This is the author’s accepted manuscript of the article published in\\ {\it BMC Medical Imaging}, 2025,  25,  Article 517, \url{https://doi.org/10.1186/s12880-025-02053-w}}


\abstract{\textbf{Background:} Gliomas are among the most common malignant brain tumors and exhibit substantial heterogeneity, complicating accurate detection and segmentation. Although multi-modal MRI is the clinical standard for glioma imaging, variability across modalities and high computational demands hamper effective automated segmentation.

\textbf{Methods:} We propose UKAN-EP, a novel 3D extension of the original 2D U-KAN model for multi-modal MRI brain tumor segmentation. While U-KAN integrates Kolmogorov-Arnold Network (KAN) layers into a U-Net backbone, UKAN-EP further incorporates Efficient Channel Attention (ECA) and Pyramid Feature Aggregation (PFA) modules to enhance inter-modality feature fusion and multi-scale feature representation. We also introduce a dynamic loss weighting strategy that adaptively balances cross-entropy and Dice losses during training.

\textbf{Results:}  On 
the 2024 BraTS-GLI dataset,
UKAN-EP achieves superior segmentation performance 
(e.g., Dice = 0.9001 $\pm$ 0.0127 
and IoU = 0.8257 $\pm$ 0.0186  for the whole tumor)
while requiring substantially fewer computational resources (223.57 GFLOPs and 11.30M parameters) compared to strong baselines including U-Net, Attention U-Net, Swin UNETR,
%{\color{black} 
VT-Unet, TransBTS,
and 3D U-KAN. An extensive ablation study further confirms the effectiveness of ECA and PFA and shows the limited utility of self-attention and spatial attention alternatives.

\textbf{Conclusion:} UKAN-EP demonstrates 
that combining the expressive power of KAN layers with  lightweight  channel-wise attention and multi-scale feature aggregation improves the accuracy and efficiency of brain tumor segmentation.
%Our findings underscore the benefit of combining interpretable function-based layers with lightweight channel-wise attention mechanisms for accurate and efficient brain tumor segmentation. 
% Code is available at \url{https://github.com/TianzeTang0504/UKAN-EP}.
}

% The abstract serves both as a general introduction to the topic and as a brief, non-technical summary of the main results and their implications. Authors are advised to check the author instructions for the journal they are submitting to for word limits and if structural elements like subheadings, citations, or equations are permitted.




\maketitle

\section{Introduction}\label{sec:introduction}

Gliomas are a prevalent form of malignant brain tumors and a leading cause of cancer-related mortality among adults \cite{price2024cbtrus}. Their invasive nature and ability to arise in any brain region  pose significant diagnostic challenges \citep{louis2020cimpact,de20242024}.
Multi-modal magnetic resonance imaging (MRI) is the gold standard for glioma imaging, providing critical insights into tumor size, location, and morphology. Commonly used MRI modalities include T1-weighted (T1), contrast-enhanced T1-weighted (T1Gd), T2-weighted (T2), and T2-weighted fluid-attenuated inversion recovery (FLAIR) \citep{verburg2021state}. Accurate segmentation of gliomas from multi-modal MRI enables precise delineation of tumor subregions, critical for comprehensive clinical assessment.

Despite its clinical importance, glioma segmentation presents several key challenges. First, gliomas exhibit substantial heterogeneity in size, location, and shape, complicating the standardization of segmentation methods \citep{visser2019inter}. Additionally, inconsistencies in intensity across MRI scans, imaging artifacts, and the need to simultaneously process and align information from multiple modalities (T1, T1Gd, T2, FLAIR) significantly increase the computational burden of accurate tumor segmentation. As high-resolution, multi-modal neuroimaging datasets become more prevalent, there is a growing need for models that can effectively integrate complementary information across modalities while maintaining high segmentation accuracy and computational efficiency.

Deep learning methods, particularly U-Net and its variants~\citep{ronneberger2015u, Ccic16, oktay2018attention, chen2021transunet, hatamizadeh2021swin}, have achieved state-of-the-art performance in medical image segmentation and have consistently ranked among the top models in recent Brain Tumor Segmentation (BraTS) challenges~\citep{myronenko20193d,jiang2020two,isensee2021nnu,hatamizadeh2021swin,ferreira2024we}. Among these, Attention U-Net~\citep{oktay2018attention} introduces attention gates to selectively enhance relevant spatial features during decoding, improving segmentation of fine structures. Swin UNETR~\citep{hatamizadeh2021swin}, on the other hand, replaces the encoder with a Swin Transformer, enabling hierarchical feature extraction through shifted window self-attention. This design captures long-range dependencies while preserving spatial resolution, making it particularly effective for complex anatomical structures.

Most deep learning architectures, including Multilayer Perceptrons (MLPs), Convolutional Neural Networks (CNNs), and transformers, rely on fixed activation functions applied at nodes such as ReLU following the linear transformation of the input. However, this design limits the model's ability to learn more flexible, interpretable nonlinear mappings~\citep{liu2024kan}. Kolmogorov-Arnold Networks (KANs)~\citep{liu2024kan} introduce a new paradigm by replacing the building blocks of MLP with learnable univariate spline functions on each edge, removing the node-based activations entirely. Thus, KANs enable learning of data-adaptive transformations directly on each connection and have shown significant improvements in both function approximation accuracy and model interpretability. To adapt  KANs for medical image segmentation, U-KAN~\citep{li2024u} incorporates Tokenized KAN blocks into the bottleneck of the U-Net architecture. By introducing KAN layers at the deepest stage where feature maps have low spatial resolution but high semantic content, U-KAN leverages the expressive capacity of KANs to better model global nonlinear relationships without disrupting the spatial priors preserved in earlier convolutional layers. This selective replacement balances efficiency, accuracy, and interpretability. Empirical results
in \citep{li2024u} show that U-KAN outperforms recent state-of-the-art models on several medical image segmentation  benchmarks, demonstrating its effectiveness in clinical scenarios where both high performance and model transparency are essential.

In this paper, we propose UKAN-EP, a novel 3D extension of the original 2D  U-KAN model~\citep{li2024u}  for multi-modal MRI brain tumor segmentation. UKAN-EP integrates Efficient Channel Attention (ECA)~\citep{wang2020eca} and Pyramid Feature Aggregation (PFA) to enhance inter-modality feature fusion and multi-scale feature representation. These components help address the challenge of effectively combining information from the T1, T1Gd, T2, and FLAIR modalities across different spatial  resolutions. We also introduce a dynamic loss weighting strategy that adaptively balances the cross-entropy and Dice losses  during training. UKAN-EP is evaluated on the BraTS-GLI dataset of the BraTS 2024 Glioma Segmentation challenge~\citep{de20242024}, and benchmarked against U-Net~\citep{Ccic16}, Attention U-Net~\citep{oktay2018attention},  Swin UNETR~\citep{hatamizadeh2021swin}, %\color{black}  
VT-Unet~\citep{VT-UNet}, and TransBTS~\citep{TransBTS}. An extensive ablation study further analyzes the contributions of ECA and PFA, and assesses the impact of modifications that utilize self-attention~\citep{vaswani2017attention} and Efficient Spatial Attention (ESA)~\citep{9460078}.

Our main contributions are summarized as follows.
%[leftmargin=1em]
\begin{itemize}
    \item  We propose UKAN-EP, a novel 3D U-Net architecture that integrates KAN, ECA, and PFA  to improve segmentation accuracy by capturing complex nonlinear patterns while maintaining computational efficiency.
    
    \item We introduce a dynamic loss weighting strategy that adaptively balances the cross-entropy and Dice losses during training.
    
    \item We evaluate UKAN-EP against leading segmentation models on the 2024 BraTS-GLI dataset and demonstrate its superior performance with significantly lower computational cost.
    
    \item We conduct an extensive ablation study to assess the individual and combined effects of ECA and PFA, and compare them against ESA and self-attention variants.
    
    \item We investigate the integration of  the Vision Transformer (ViT) block \citep{dosovitskiy2020vit} into the U-KAN architecture and find that it provides no performance gains while introducing training instability.

\end{itemize}

The rest of the paper is organized as follows.
Section~\ref{sec: methods} introduces the  UKAN-EP network architecture.
Section~\ref{sec:experiments} describes the 2024 BraTS-GLI dataset,
evaluation metrics, and training details.
Section~\ref{sec:results} presents the results on segmentation performance, ablation study, and computational efficiency. 
%{\color{black} 
Section~\ref{sec:discussion} discusses the advances of UKAN-EP over existing models,  as well as its limitations and potential directions for future work.
Section~\ref{sec: conclusion} concludes the paper.
Code is available at \url{https://github.com/TianzeTang0504/UKAN-EP}.

\section{Method}\label{sec: methods}
\subsection{U-KAN (U-Net based on Kolmogorov-Arnold Network)} %{\color{black}

\subsubsection{Kolmogorov-Arnold Network (KAN)} %{\color{black}

%\subsection{Kolmogorov-Arnold Network (KAN)}
KAN \citep{liu2024kan}  is inspired by the
 Kolmogorov-Arnold representation theorem \citep{kolmogorov:superposition}, which  states that any multivariate continuous function $f:[0,1]^d \to \mathbb{R}$ can be written with
 univariate continuous functions
$\{\psi_i,\phi_{ij}\}$
as
\[
f(x_1,\dots,x_d)=\sum_{i=1}^{2d+1}\psi_i\Big(\sum_{j=1}^d\phi_{ij}(x_j)\Big).
\]
This result  naturally suggests a two-layer neural network structure: in the inner layer, each input variable undergoes univariate nonlinear transformations $\{\phi_{ij}\}$ to extract local features;
in the outer layer, these transformed features are linearly combined and  passed through another set of univariate nonlinear transformations $\{\psi_i\}$ to generate global features, which are subsequently aggregated to produce the final output.

KAN practically generalizes this two-layer neural network to arbitrary widths and depths, fitting the univariate functions using B-splines. Specifically,
let $\mathbf{\Phi}_k=[\phi_{k,i,j}(\cdot)]_{1\le i\le d_k,1\le j\le d_{k-1}}$ be the function matrix corresponding to the $k$-th KAN layer, and define 
$$\mathbf{\Phi}_k(\mathbf{v})=
\left(\sum_{j=1}^{d_{k-1}}\phi_{k,1,j}(v_j),\dots, \sum_{j=1}^{d_{k-1}}\phi_{k,d_k,j}(v_j)\right)^\top
~~\text{for}~~\mathbf{v}=(v_1,\dots,v_{d_{k-1}})^\top.
$$
For an input $\mathbf{x}\in\mathbb{R}^{d_0}$, a $K$-layer KAN is then given by
\begin{align*}
\text{KAN}(\mathbf{x})
=\mathbf{\Phi}_K \circ \mathbf{\Phi}_{K-1}
\circ \cdots \circ  \mathbf{\Phi}_1(\mathbf{x}).
\end{align*}
Each univariate function $\phi_{k,i,j}$  is parameterized as a B-spline curve,
whose parameters are learned during training. 
In contrast, 
based on the universal representation theorem \citep{hornik1989multilayer},
MLP is written as
$$
\text{MLP}(\mathbf{x})=\mathbf{W}_K\circ \sigma \circ\mathbf{W}_{K-1}\circ \sigma\circ\cdots \circ
\mathbf{W}_2 \circ \sigma \circ \mathbf{W}_1(\mathbf{x}),
$$
where each $\mathbf{W}_k$ is an affine transformation  with trainable weight and bias parameters, and $\sigma$ is a fixed nonlinear activation function. Structurally, MLP uses the same fixed function $\sigma$ on nodes, whereas KAN substitutes learnable activation functions $\{\phi_{k,i,j}\}$ for the weight parameters of $\{\mathbf{W}_k\}$ on edges.
Therefore, KAN offers enhanced interpretability, while often achieving comparable or superior performance to MLP with significantly fewer trainable parameters.

\subsubsection{U-KAN}\label{sec:method_ukan} %{\color{black}
% \noindent\textbf{U-KAN.}

The U-KAN model~\citep{li2024u} integrates KAN layers~\citep{liu2024kan} into the traditional U-Net structure~\citep{ronneberger2015u}. 
%In our experiment, we use the default U-KAN configuration, with channel sizes \( D_1 = 128 \), \( D_2 = 160 \), and \( D_3 = 256 \) (see Figure~\ref{fig:ukan}). 
The network architecture employs a two-phase design: a convolution phase for initial feature extraction, followed by a Tokenized KAN (Tok-KAN) phase where the KAN layers refine the feature representations. Specifically, the KAN layers in the Tok-KAN phase process tokenized features using B-spline based activation functions to model complex patterns. For an input feature tensor 
$\mathbf{X}_{k-1}$, the $k$-th  Tok-KAN block is formulated~as
\begin{align*}
    \mathbf{X}_{k} = \mathrm{LayerNorm}(\mathbf{X}_{k-1} + \mathrm{DwConv}(\mathrm{KAN}(\mathrm{Tok}(\mathbf{X}_{k-1})))),
\end{align*}
where $\mathrm{KAN}(\mathrm{Tok}(\mathbf{X}_{k-1}))$ applies the KAN layer to the tokenized features 
$\mathrm{Tok}(\mathbf{X}_{k-1})$ with learnable activation functions, followed by depth-wise convolution (DwConv), layer normalization (LayerNorm), and residual connection for stability.

It is important to note that the original U-KAN was designed for 2D image segmentation.
Motivated by its strong performance on 2D tasks, we hypothesize that U-KAN can also perform well in 3D applications such as brain tumor segmentation.
To this end, we adapt the model to a 3D version by replacing all 2D operations (e.g., 2D convolutions) with their 3D counterparts.
Notably, the Tok-KAN block does not impose fixed spatial or dimensional constraints on its input, as all feature maps are tokenized via patching, vectorization, and a convolutional layer.

\vspace{2em}
\subsection{Efficient Channel Attention (ECA)}
\label{sec:eca}

ECA~\citep{wang2020eca} is a lightweight channel attention mechanism that enhances feature representation without significantly increasing computational complexity. Traditional channel attention mechanisms \citep{hu2018squeeze,woo2018cbam} employ fully connected layers to capture cross-channel interactions, necessitating channel dimensionality reduction to manage model complexity, which can adversely affect the learning of channel attention. In contrast, ECA efficiently learns  channel attention
by modeling cross-channel interactions using a simple 1D convolution
without dimensionality reduction.
The ECA process consists of three key steps:
\begin{enumerate}[leftmargin=1.7em]
    \item \textbf{Global Feature Compression:}  
    Global average pooling (GAP) is applied to the spatial dimensions of the input feature tensor \( \mathbf{X}=[X_{c,d,h,w}] \in \mathbb{R}^{C \times D\times H \times W } \), resulting in aggregated features \( \mathbf{z}=(z_1,\dots, z_C) \in \mathbb{R}^C \):  
    \[
    z_c = \frac{1}{D\times H \times W } \sum_{d=1}^D\sum_{h=1}^{H} \sum_{w=1}^{W} X_{c, d, h, w} \quad
 \text{for}   
    \quad c = 1, 2, \ldots, C.
    \]

    \item \textbf{Local Cross-Channel Interaction Modeling:}  
    A 1D convolution  is applied to \( \mathbf{z} \) to capture local interactions among channels,
    followed by a sigmoid function
    to generate the channel weights:
    \[
(a_1,\dots,a_C) = \text{Sigmoid(Conv1D}(\mathbf{z}, k)).
    \]
 %   where \( \mathbf{a} \in \mathbb{R}^C \) contains the attention weights. 
    %The kernel size \( k \) is set to 3 in our experiment.
    %adaptively determined based on the number of channels \( C \), ensuring symmetry by making \( k \) an odd number.  

    \item \textbf{Feature Recalibration:}  
    The channel weights $(a_1,\dots,a_C)$ are applied to each channel of the input feature tensor $\mathbf{X}=[\mathbf{X}_1;\dots;\mathbf{X}_C]$
 to produce
 a recalibrated feature tensor
 $\tilde{\mathbf{X}}=[\tilde{\mathbf{X}}_1;\dots;\tilde{\mathbf{X}}_C]$, where  
 informative feature channels
 are emphasized
 and less useful ones
 are suppressed: 
    \[
    \tilde{\mathbf{X}}_c = a_c  \mathbf{X}_c
    \quad
 \text{for}   
    \quad
    c = 1, 2, \ldots, C.
    \]
\end{enumerate}

%ECA’s simplicity and effectiveness make it ideal for reducing computational complexity while significantly enhancing the network's ability to focus on important features.

\subsection{Pyramid Feature Aggregation (PFA)}
\label{sec:method_pfa}
Merging semantically rich deep features with spatially precise shallow features is a common strategy in hierarchical fusion frameworks~\citep{lin2017feature, zhang2021pyramid, zhou2018unet++}. Building on this principle, we introduce a Pyramid Feature Aggregation (PFA) module to enhance multi-scale representation. Let \( \{ \mathbf{X}^{(l)} \}_{l=1}^3 \) denote the encoder feature maps from shallowest (\( l = 1 \)) to deepest (\( l = 3 \)). The PFA module proceeds in a top-down manner from deep to shallow. At each stage \( l\in\{1,2\}\), the upsampled output from the deeper recalibrated feature \( \tilde{\mathbf{X}}^{(l+1)} \)  (with $\tilde{\mathbf{X}}^{(3)}=\mathbf{X}^{(3)}$) is concatenated with the current encoder feature \( \mathbf{X}^{(l)} \):
\[
\check{\mathbf{X}}^{(l)} = \text{Concat}(\text{Upsample}(\tilde{\mathbf{X}}^{(l+1)}), \mathbf{X}^{(l)}).
\]
The aggregated tensor \( \check{\mathbf{X}}^{(l)} \) is then passed through the ECA module (Section~\ref{sec:eca}) to produce the recalibrated output \( \tilde{\mathbf{X}}^{(l)} \). The final outputs \( \{ \tilde{\mathbf{X}}^{(l)} \}_{l=1}^{2} \) are propagated as skip connections to the decoder. This structure facilitates hierarchical fusion, enhancing cross-scale feature continuity and improving segmentation precision compared to conventional U-Net designs.

\begin{figure}[t!]
    \centering
    \includegraphics[width=\textwidth]{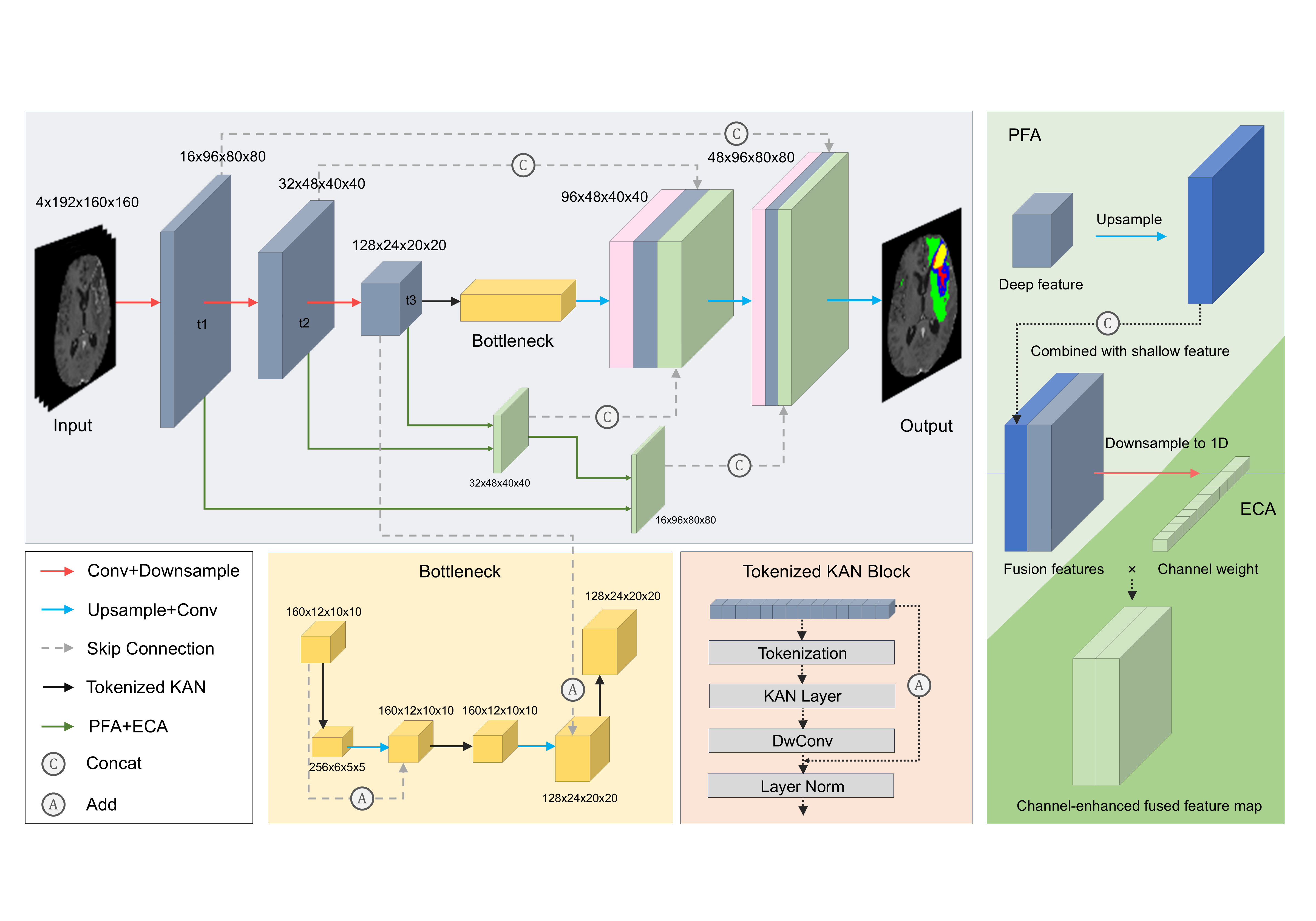}
\caption{Architecture of UKAN-EP. The model  combines Tokenized KAN blocks at the bottleneck with Pyramid Feature Aggregation (PFA) and Efficient Channel Attention (ECA) modules to achieve enhanced integration and refinement of multi-modal features.}
    \label{fig:ukan}
\end{figure}

\subsection{The Proposed UKAN-EP}
As illustrated in Figure~\ref{fig:ukan}, UKAN-EP extends the original U-KAN architecture~\citep{li2024u}, a U-Net variant that incorporates Tok-KAN blocks at the bottleneck. Let \( \mathbf{X}^{(1)}, \mathbf{X}^{(2)}, \mathbf{X}^{(3)} \) denote the encoder outputs from shallowest to deepest layers (corresponding to \( t1, t2, t3 \) in the figure). These multi-scale features are progressively fused using the PFA modules, which upsample deeper features and concatenate them with shallower ones (Section~\ref{sec:method_pfa}). The fused tensors are then recalibrated using the ECA modules, which apply lightweight 1D convolutions to capture channel-wise dependencies without dimensionality reduction (Section~\ref{sec:eca}). The deepest encoder feature map, $\mathbf{X}^{(3)}$, is passed to the Tok-KAN blocks in the bottleneck, where it is tokenized, processed by spline-based KAN layers to capture nonlinear interactions, and restored to spatial format via depth-wise convolution and layer normalization (Section~\ref{sec:method_ukan}). This replaces traditional MLPs with interpretable univariate function compositions, enhancing transparency in representation learning for high-level features. The outputs from the PFA+ECA blocks serve as additional skip connections to the decoder and are concatenated with the upsampled decoder outputs at each resolution level. The effectiveness of this design is  confirmed by the ablation study  in Section~\ref{sec:result_ablation}. 

%{\color{black} 
A key advantage of UKAN-EP is the interpretability of its Tok-KAN blocks. Unlike standard MLPs that use fixed nonlinear activations, KAN layers employ spline-based learnable activation functions on edges, enabling direct visualization of the learned mappings. This property allows us to qualitatively inspect how different univariate functions are adapted during training. In our experiments, the learned KAN functions exhibit smooth and localized variations, revealing how the model selectively amplifies or suppresses feature responses. This yields transparent, function-level views of the bottleneck transformations—an interpretability property that conventional U-Net and most transformer encoders do not provide natively.

\subsection{Loss Function}\label{sec: loss function}
We adopt a dynamic weighting strategy to combine the cross-entropy loss~\citep{zhang2018generalized} and the Dice loss~\citep{sudre2017generalised}.
The total loss is defined as 
\[
\mathcal{L}_{\text{total}} =\frac{1}{B}\sum_{i=1}^B \left\{ (1 - \alpha_i) \cdot \mathcal{L}_{\text{CE}}^{(i)} + \alpha_i \cdot \mathcal{L}_{\text{Dice}}^{(i)}\right\}, 
\]
% where
% \begin{align}
% \alpha_i &= \frac{\mathcal{L}_{\text{CE}}^{(i)}}{\mathcal{L}_{\text{CE}}^{(i)} + \mathcal{L}_{\text{Dice}}^{(i)}}, \nonumber\\
% \mathcal{L}_{\text{CE}}^{(i)} &= - \sum_{v=1}^{N} \sum_{c=1}^{C} y_{v,c}^{(i)} \log \hat{y}_{v,c}^{(i)}, \nonumber\\
% \mathcal{L}_{\text{Dice}}^{(i)} &= 1 - \frac{2 \sum_{v=1}^{N} \sum_{c=2}^{C} \hat{y}_{v,c}^{(i)} \cdot y_{v,c}^{(i)}}{\sum_{v=1}^{N} \sum_{c=2}^{C} \hat{y}_{v,c}^{(i)} + \sum_{v=1}^{N} \sum_{c=2}^{C} y_{v,c}^{(i)}}, \label{individual dice loss}
% \end{align}

where
\begin{align}
\alpha_i &= \frac{\mathcal{L}_{\text{CE}}^{(i)}}{\mathcal{L}_{\text{CE}}^{(i)} + \mathcal{L}_{\text{Dice}}^{(i)}}, \nonumber\\
\mathcal{L}_{\text{CE}}^{(i)} &= - {\color{black}\frac{1}{N}} \sum_{v=1}^{N} \sum_{c=1}^{C} y_{v,c}^{(i)} \log \hat{y}_{v,c}^{(i)}, \nonumber\\
\mathcal{L}_{\text{Dice}}^{(i)} &= 1 - \frac{2 \sum_{v=1}^{N} \sum_{c=2}^{C} \hat{y}_{v,c}^{(i)}\, y_{v,c}^{(i)} }
{\sum_{v=1}^{N} \sum_{c=2}^{C} \hat{y}_{v,c}^{(i)} + \sum_{v=1}^{N} \sum_{c=2}^{C} y_{v,c}^{(i)} 
{\color{black} + \epsilon}}, \label{individual dice loss}
\end{align}
\( y_{v,c}^{(i)} \in \{0,1\} \) 
is the one-hot ground-truth indicator that voxel \( v \in \{1, \ldots, N\} \) 
in image $i\in\{1,\dots,B\}$ belongs to 
class \( c \in \{1, \ldots, C\} \),  \( \hat{y}_{v,c}^{(i)} \in [0,1] \) is the corresponding predicted softmax probability,
{\color{black}
and $\epsilon=10^{-6}$
is added to prevent division by zero.}
{\color{black}The background class \( c = 1 \) is excluded from the Dice loss to focus on foreground regions.} 
This formulation integrates the voxel-wise classification strength of cross-entropy loss with the overlap-based sensitivity of Dice loss, enabling more complementary learning. 
The dynamic coefficient \( \alpha_i \in (0,1) \) is updated at each iteration based on the relative magnitude of the cross-entropy and Dice losses, ensuring that neither dominates the training. We show in Section~\ref{sec:abl_loss} that dynamic  weighting enhances segmentation performance compared to applying fixed weights. 

\section{Experiments}\label{sec:experiments}

\subsection{Data Description}\label{sec:Data Description}
We use the 2024 BraTS-GLI dataset, a multi-modal MRI dataset provided in Task~1 of the 2024 BraTS Challenge~\citep{de20242024}
(\url{https://www.synapse.org/Synapse:syn53708249/wiki/627759}), which focuses on automated segmentation of post-treatment glioma subregions in adults. This dataset is part of the annual MICCAI BraTS challenge~\citep{menze2014multimodal,bakas_advancing_2017,baid2021rsnaasnrmiccai}, which aims to benchmark methods for delineating tumor structures from clinical multi-parametric MRI  scans. 
%{\color{black}To assess model robustness and generalizability beyond the training distribution, we additionally evaluate performance on the BraTS 2025 external dataset.} 
All MRI scans were acquired from multiple academic medical centers and preprocessed following a standardized pipeline consistent with the 2017–2023 BraTS challenges~\citep{de20242024}. Raw DICOM-format scans were first reviewed by institutional radiologists, after which T1, T1Gd, T2, and FLAIR sequences were extracted and renamed according to the BraTS naming convention. The scans were then converted to NIfTI format using the \texttt{dcm2niix} tool~\citep{nifti}. Brain extraction was performed using HD-BET~\citep{isensee2019automated} to remove non-brain tissue (e.g., neck fat, skull, eyeballs). All sequences were subsequently co-registered to the Linear Symmetrical MNI atlas using affine registration via CapTK/Greedy~\citep{pati2020cancer}. The final preprocessed volumes have dimensions of $218 \times 182 \times 182$ voxels per modality.

\begin{figure}[t!]
    \centering
    \includegraphics[width=\textwidth]{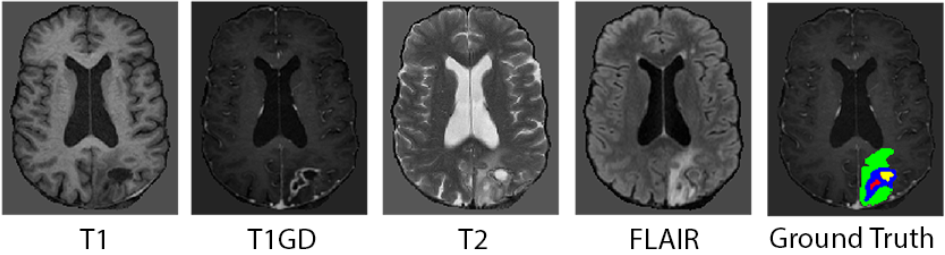}
    \caption{Example slices from the four MRI modalities and the ground-truth segmentation. For the truth labels, red is NETC, green is SNFH, blue is ET, and yellow is RC.}
    \label{fig:enter-label}
\end{figure}

Each subject has four 3D MRI modalities
including T1, T1Gd, T2, and FLAIR,
as illustrated in Figure~\ref{fig:enter-label}. These modalities provide complementary anatomical and pathological information: T1 offers structural detail, T1Gd highlights enhancing tumor regions, T2 captures edema, and FLAIR visualizes periventricular signal abnormalities by suppressing cerebrospinal fluid. The ground truth segmentations define four primary tumor subregions: enhancing tissue (ET), non-enhancing tumor core (NETC), surrounding non-enhancing FLAIR hyperintensity (SNFH), and resection cavity (RC). A fifth composite label, whole tumor (WT), is defined as the union of ET, NETC, and SNFH, and serves as an aggregate measure for overall segmentation performance. ET captures regions of active tumor and nodular enhancement; NETC denotes necrotic or cystic components within the tumor; SNFH includes edema, infiltrative tumor, and post-treatment signal abnormalities; and RC encompasses recent or chronic surgical cavities typically containing fluid, blood, or other proteinaceous materials~\citep{de20242024}. 
%{\color{black}
The challenge dataset, provided during the 2024 BraTS Challenge, consists of 1350 labeled post-treatment glioma cases and 188 unlabeled cases. For model development and evaluation, the 1350 labeled cases are randomly split into 1080 training, 135 validation, and 135 test samples using an 8:1:1 ratio. \cynthia{In addition, a post-challenge dataset, released after the 2024 BraTS Challenge and made available on the official website, contains 271 newly labeled post-treatment cases and serves as an additional test set.}

\subsection{Data Processing and Augmentation}\label{sec:data_aug}

Each subject's MRI scans consisting of four modalities (T1, T1Gd, T2, and FLAIR) are combined into a single 4D volume of shape \(C \times D \times H \times W \), where \(C = 4\). Non-brain voxels are masked to zero to suppress irrelevant intensity variation. During training, a series of online data augmentation techniques is applied to enhance model generalization and robustness to acquisition variability. A crop of size \(4 \times 192 \times 160 \times 160\) is first extracted to retain the brain  while reducing computational load. Random flipping is then performed independently along each anatomical axis (\(x\), \(y\), \(z\)) with a Bernoulli probability of 0.5, promoting spatial invariance. To simulate variability in image quality, Gaussian noise sampled from \(\mathcal{N}(0, 0.01^2)\) is added to non-background voxels. Spatial misalignment is addressed by applying random rotations uniformly sampled from \([-10^{\circ}, 10^{\circ}]\) around arbitrarily selected axis pairs. Images are resampled using trilinear interpolation, while segmentation label maps are assigned via nearest neighbor interpolation. Finally, random contrast scaling is applied with multiplicative factors drawn from the uniform distribution \(\mathcal{U}(0.8, 1.2)\), preserving the mean intensity while modulating contrast distribution.

\subsection{Evaluation Metrics}\label{sec:evaluation metrics}
Segmentation performance is evaluated using three standard metrics \citep{taha2015metrics}: Dice similarity coefficient (Dice), Intersection over Union (IoU), and the 95th percentile Hausdorff Distance (HD95). Let \(P\) and \(G\) denote the predicted and ground truth segmentation masks, respectively. Dice and IoU evaluate the degree of volumetric overlap between \(P\) and \(G\), while HD95 quantifies the spatial deviation between boundaries of \(P\) and \(G\). All metrics are computed for each of the five tumor subregions (ET, NETC, SNFH, RC, and WT) to enable detailed evaluation of segmentation performance across clinically relevant compartments.

Dice is the degree of overlap between \(P\) and \(G\), defined as
\[
    \text{Dice}(P, G) = \frac{2 |P \cap G|}{|P| + |G|}.
\]
Dice ranges from 0 to 1, with higher values indicating greater agreement. By focusing on foreground overlap rather than background agreement, Dice is particularly effective for evaluating segmentation performance in class imbalanced settings.

IoU, also known as the Jaccard index, is defined as
\[
    \text{IoU}(P, G) = \frac{|P \cap G|}{|P \cup G|}.
\]
IoU also ranges from 0 to 1 and, in contrast to Dice, assigns proportionally more weight to false positives and false negatives relative to true positives, making it more sensitive to misclassification and a stricter metric for overlap quality.

HD95 is a robust metric for evaluating the
spatial alignment between the predicted and ground truth segmentation boundaries. Let \( d(x, A) = \inf_{a \in \partial A} \|x - a\| \) denote the shortest Euclidean distance from a point \(x\) to  the boundary $\partial A$ of set \(A\).  HD95 is defined as
\[
\text{HD95}(P, G) = \operatorname{percentile}_{95} \left(
\{ d(p, G) : p \in \partial P \} \cup \{ d(g, P) : g \in \partial G \}
\right).
\]
It computes the 95th percentile
of
the distances between
the closest points of the two boundaries
to reduce sensitivity to outliers, providing a symmetric and robust estimate of boundary error. A HD95 value of 0 indicates perfect boundary alignment.

\subsection{Training Details}\label{sec:hyperparameter}
All models are implemented in PyTorch and trained on an NVIDIA RTX 8000 GPU  (48GB VRAM) with an Intel Xeon Gold 6244 CPU (8 cores, 3.6GHz, 200GB RAM). Input volumes are cropped  to a size of  \(4\times 192 \times 160 \times 160\)
to retain the brain, with four MRI modalities concatenated along the channel axis, as described in Section~\ref{sec:data_aug}. Each  model is trained for 300 epochs using a batch size of 2, which accommodates the high memory demands of 3D MRI volumes.
The AdamW optimizer~\citep{loshchilov2017decoupled} is used with a weight decay of $0.0001$. 
Learning rates are scheduled using cosine annealing~\citep{loshchilov2016sgdr}. For most models, the schedule starts at 0.005, peaks at 0.01 after 30 warm-up epochs, and decays gradually over the remaining epochs. 
Swin UNETR follows a similar schedule but uses a smaller initial learning rate of 0.001, peaking at 0.005 after 30 warm-up epochs before decaying.
These learning rate settings reflect empirically tuned values that yielded stable and strong performance for each model. %{\color{black} 
All model weights are initialized using standard  schemes to ensure stable convergence during 3D training. Linear layers are initialized with a truncated normal distribution with mean~0, standard deviation~0.02, and values truncated to the range 
$[-0.04,0.04]$. Convolutional layers are initialized with Kaiming normal initialization, and layer normalization weights and biases are initialized to 1 and 0, respectively.
%}

%{\color{black}
\section{Results}\label{sec:results}
%\section{Results and Discussion}\label{sec:results}

This section presents a comprehensive evaluation of the proposed UKAN-EP model. In Section~\ref{sec:result_seg_perf}, the segmentation performance across the five tumor subregions is discussed. In Section~\ref{sec:result_ablation}, an extensive ablation study is provided, assessing the impact of PFA, ECA, and additional attention-based architectural modifications. Finally, in Section~\ref{sec:result_comp}, the computational efficiency of UKAN-EP is presented.
\vspace{1em}
\subsection{Segmentation Performance}\label{sec:result_seg_perf}

We evaluate %{\color{black} 
our proposed UKAN-EP against  six models: classical U-Net~\citep{Ccic16}, Attention U-Net (Att-Unet)~\citep{oktay2018attention}, Swin UNETR~\citep{hatamizadeh2021swin}, %{\color{black} 
VT-Unet~\citep{VT-UNet}, TransBTS~\citep{TransBTS}, and 
U-KAN~\citep{li2024u}. U-Net captures multi-scale features via its encoder-decoder structure, while Att-Unet augments this design with attention gates to improve focus on relevant regions. Swin UNETR combines a U-shaped architecture with a Swin Transformer encoder for hierarchical attention. \cynthia{VT-UNet is a fully transformer-based architecture, and TransBTS integrates 3D CNN encoders with transformer layers.} U-KAN introduces spline-based KAN layers into the U-Net framework as described in Section~\ref{sec:method_ukan}.
We use the 3D version of U-KAN,
which corresponds to our UKAN-EP model  (Figure~\ref{fig:ukan}) without ECA and PFA modules.

Table~\ref{tab:dice} reports the  segmentation performance in terms of Dice, IoU, and HD95 
averaged over %{\color{black} 
135 test cases from the challenge dataset 
for each of the five tumor subregions: 
ET, NETC, SNFH, RC, and WT. UKAN-EP displays the highest volumetric overlap with the ground truth in four  of the five subregions, achieving the highest average Dice scores of 0.5197, 0.8887, 0.6924, and 0.9001, and the highest average IoU scores of 0.4238, 0.8086, 0.6156, 0.8257 for NETC, SNFH, RC, and WT, respectively. For the segmentation of the ET region, Swin UNETR performs best, with an average Dice score 0.0027 and average IoU 0.0049  higher than those of UKAN-EP. Best boundary alignment for NETC and RC is achieved by UKAN-EP. Att-Unet yields the lowest average HD95 for SNFH, and U-Net attains the lowest average HD95 for ET and WT. 
The uncertainties of these average  metric values are comparably stable across all \cynthia{seven} methods as shown in Table~\ref{tab:SD}. We notice a significant improvement in performance by the addition of PFA and ECA modules to U-KAN, highlighting the advantage of multi-scale feature aggregation and channel-wise recalibration. This is further examined in the ablation study in Section~\ref{sec:result_ablation}. 

%{\color{black}Notably, only the proposed UKAN-EP avoids the false positives of SNFH (in green) in the upper-left region of the slice.}

%{\color{black} 
Table~\ref{tab:dice_271} presents the average Dice, IoU, and HD95 metrics on the 271 post-challenge test  cases,
with the corresponding uncertainties reported in Table~\ref{tab:se_271}. UKAN-EP maintains strong segmentation performance and generalizes well to this additional test set, outperforming other U-Net variants in most evaluation categories. The reported uncertainties are stable across all seven methods. These results further support the robustness and generalizability of the proposed method.
%}

\begin{figure}[h!]
    \centering
    \includegraphics[width=0.72\textwidth]{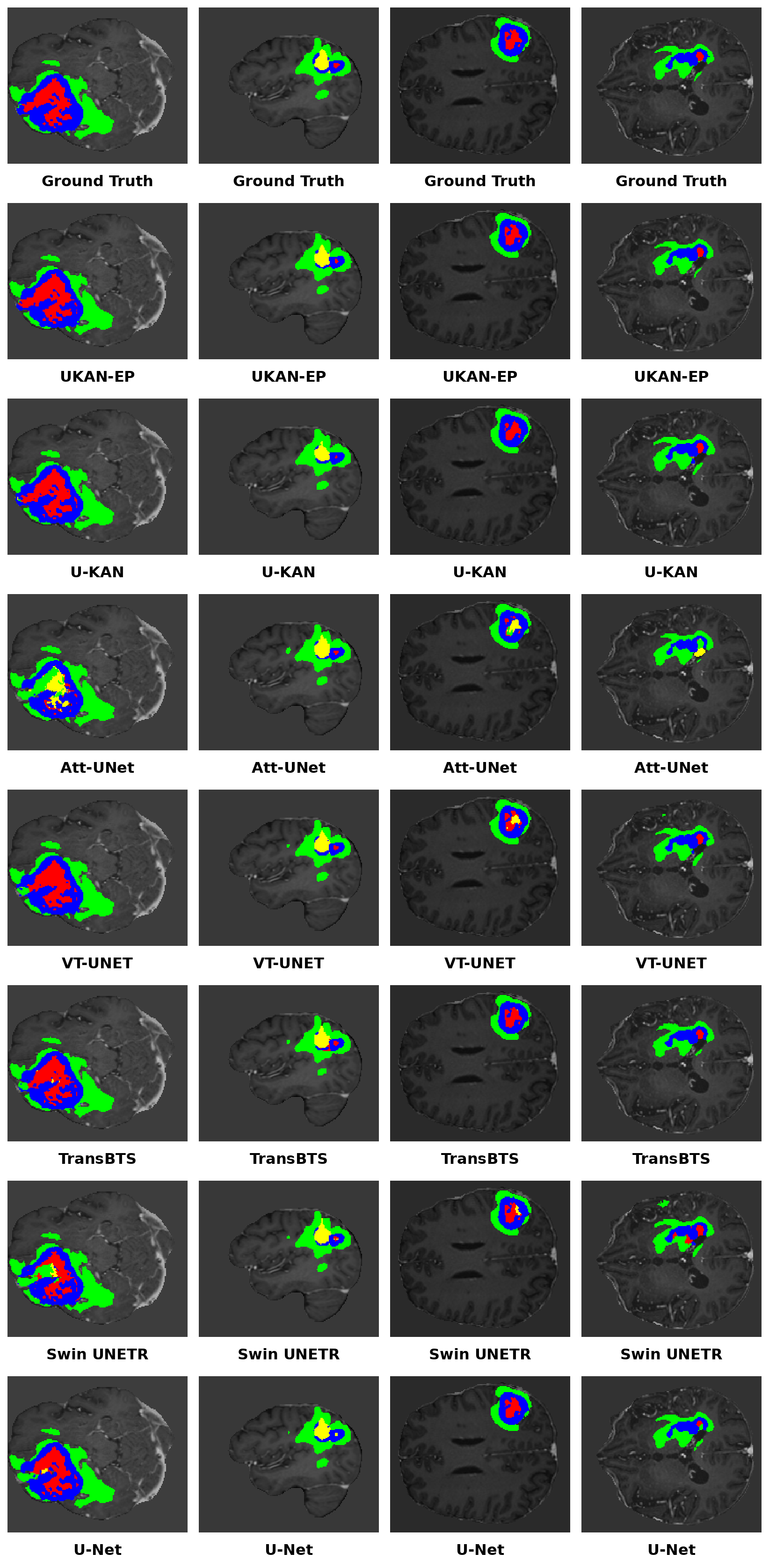}
\caption{Example segmentation results for 
four BraTS-GLI test cases from the 2024    BraTS Challenge dataset (first two columns) and the post-Challenge dataset (last two columns).
 Red is NETC, green is SNFH, blue is ET, and yellow is RC.}    \label{fig:finalseg}
\end{figure}

The segmentation results are illustrated in Figure~\ref{fig:finalseg}. This figure presents representative examples from four test cases, including two from the challenge dataset and two from the post-challenge dataset.
The visual comparison demonstrates that UKAN-EP achieves more complete and smoother tumor delineations across the NETC, SNFH, ET, and RC  regions, exhibiting superior boundary continuity and structural consistency.
Swin UNETR tends to produce fragmented segmentation results, likely due to its heavy encoder design, which leads to overfitting during feature extraction and limited generalization on small 3D datasets. Att-Unet shows similar behavior, especially in the NETC region, where its attention mechanism may drift
under complex lesion morphology or low image contrast, resulting in incomplete localization.
U-Net and TransBTS demonstrate suboptimal performance in fine boundary delineation, possibly because their feature fusion stages oversmooth high-frequency spatial details, reducing sensitivity to subtle structural variations. VT-UNet exhibits a high misclassification rate, indicating that while its global modeling capability is strong, it lacks sufficient local contextual constraints.
Although U-KAN and UKAN-EP perform comparably overall, the former shows slightly inferior segmentation and produces more misclassifications.
In contrast, UKAN-EP generates sharper lesion boundaries and effectively suppresses spurious predictions, highlighting the contributions of the PFA module in maintaining spatial consistency and  the ECA module in refining feature weighting and channel selection, thereby achieving a balanced integration of global perception and local discrimination.
These qualitative observations are consistent with the quantitative improvements reported in Tables \ref{tab:dice}–\ref{tab:se_271}, confirming the robustness and boundary precision of the proposed UKAN-EP.

\begin{sidewaystable}[t!]
\setlength{\tabcolsep}{2pt}
\caption{Average evaluation metrics on the test set (135 cases). 
ET = enhancing tissue, NETC = non-enhancing tumor core, RC = resection cavity, 
SNFH = surrounding non-enhancing FLAIR hyperintensity, WT = ET+SNFH+NETC.}\label{tab:dice}
%\centering
%\scriptsize
\tiny
\begin{tabular}{l|ccccc|ccccc|ccccc} 
\hline
 & \multicolumn{5}{c|}{\underline{Dice}} & \multicolumn{5}{c|}{\underline{IoU}} & \multicolumn{5}{c}{\underline{HD95}} \\ 
Model & {NETC} & {SNFH} & {ET} & {RC} & {WT} 
  & {NETC} & {SNFH} & {ET} & {RC} & {WT} 
  & {NETC} & {SNFH} & {ET} & {RC} & {WT}  \\ \hline
{U-Net} &0.4477 & 0.8858 & 0.6001 & 0.6724 & 0.8962& 0.3621& 0.8049& 0.5118& 0.5938 & 0.8204& 4.8885 & 3.1117 & \textbf{4.5522} & 7.7096& \textbf{3.0945}\\
{Att-Unet} &0.3572 & 0.8879 & 0.5811 & 0.6577 & 0.8979& 0.2673 & 0.8082 & 0.4938 & 0.5811 & 0.8228& 3.5663 & \textbf{2.9797} & 4.9709& 8.1658& 3.1305\\
{Swin UNETR} &0.4374& 0.8800& \textbf{0.6176}& 0.6454& 0.8912& 0.3553& 0.7981& \textbf{0.5327}& 0.5685& 0.8144& 4.5571& 3.7400& 6.1675& 7.2218& 3.8368\\
 {VT-Unet}& 0.5137& 0.8869& 0.6148& 0.6391& 0.8892& 0.4126& 0.7864& 0.5153& 0.5819& 0.8154& 3.6567& 4.7233& 5.2935& 6.3971&3.7310\\
 {TransBTS}& 0.4672& 0.8638& 0.5842& 0.6264& 0.8777& 0.3800& 0.7741& 0.4902& 0.5400& 0.7936& 3.9009& 3.4054& 5.4997& 8.1699&3.4817\\
 
% \textcolor{blue}{{VT-Unet}} & \textcolor{blue}{0.5137} & \textcolor{blue}{0.8869} & \textcolor{blue}{0.6148} & \textcolor{blue}{0.6391} & \textcolor{blue}{0.8892} &
% \textcolor{blue}{0.4126} & \textcolor{blue}{0.7864} & \textcolor{blue}{0.5153} & \textcolor{blue}{0.5819} & \textcolor{blue}{0.8154} &
% \textcolor{blue}{3.6567} & \textcolor{blue}{4.7233} & \textcolor{blue}{5.2935} & \textcolor{blue}{6.3971} & \textcolor{blue}{3.7310} \\

% \textcolor{blue}{{TransBTS}} & \textcolor{blue}{0.4672} & \textcolor{blue}{0.8638} & \textcolor{blue}{0.5842} & \textcolor{blue}{0.6264} & \textcolor{blue}{0.8777} &
% \textcolor{blue}{0.3800} & \textcolor{blue}{0.7741} & \textcolor{blue}{0.4902} & \textcolor{blue}{0.5400} & \textcolor{blue}{0.7936} &
% \textcolor{blue}{3.9009} & \textcolor{blue}{3.4054} & \textcolor{blue}{5.4997} & \textcolor{blue}{8.1699} & \textcolor{blue}{3.4817} \\
{U-KAN} &0.4526& 0.8810& 0.6152& 0.6860& 0.8927& 0.3589& 0.7996& 0.5200& 0.6079& 0.8169& 3.1712& 3.6289& 5.7397& 6.3909& 4.5206\\
 {UKAN-EP} & \textbf{0.5197}& \textbf{0.8887}& 0.6149& \textbf{0.6924}& \textbf{0.9001}& \textbf{0.4238}&\textbf{0.8086}&0.5278&\textbf{0.6156}&\textbf{0.8257}&\textbf{2.7882}&3.1934&4.8962&\textbf{4.4771}&3.1425\\
\hline
\end{tabular}
%\end{sidewaystable}

\bigskip
\bigskip
\bigskip

%\begin{sidewaystable}[h!]
\setlength{\tabcolsep}{2pt}
\caption{Uncertainties (1.96 standard errors) of the average evaluation metrics  on the test set (135 cases).}\label{tab:SD}
\centering
\tiny
\begin{tabular}{l|ccccc|ccccc|ccccc}
\hline
& \multicolumn{5}{c|}{\underline{Dice}} & \multicolumn{5}{c|}{\underline{IoU}} & \multicolumn{5}{c}{\underline{HD95}} \\ 
Model & {NETC} & {SNFH} & {ET} & {RC} & {WT}
 & {NETC} & {SNFH} & {ET} & {RC} & {WT}
 & {NETC} & {SNFH} & {ET} & {RC} & {WT}  \\ \hline
{U-Net}
& 0.0835 & 0.0153 & 0.0641 & 0.0612 & 0.0139 
& 0.0739 & 0.0208 & 0.0597 & 0.0583 & 0.0195
& 1.6973 & 0.7638 & 1.3797 & 2.4301 & 0.7349 \\
{Att-Unet}
& 0.0730 & 0.0152 & 0.0645 & 0.0619 & 0.0135
& 0.0614 & 0.0209 & 0.0605 & 0.0590 & 0.0192
& 1.1183 & 0.6477 & 1.3839 & 2.2422 & 0.6321 \\
{Swin UNETR}& 0.0821 & 0.0174 & 0.0630 & 0.0619 & 0.0159
& 0.0715 & 0.0229 & 0.0603 & 0.0590 & 0.0213
& 1.5832 & 1.0443 & 2.0758 & 2.0196 & 1.0425 \\
 {VT-Unet}& 0.0892& 0.0146& 0.0649& 0.0606& 0.0128& 0.0822& 0.0203& 0.0621& 0.0589& 0.0185& 1.3986& 0.7673& 1.6730& 2.1473&0.7535\\
 {TransBTS}& 0.0825& 0.0187& 0.0621& 0.0607& 0.0168& 0.0734& 0.0243& 0.0578& 0.0568& 0.0224& 1.2659& 0.7141& 1.4367& 2.1414&0.6793\\
 
% \textcolor{blue}{{VT-Unet}} & \textcolor{blue}{0.0892} & \textcolor{blue}{0.0146} & \textcolor{blue}{0.0649} & \textcolor{blue}{0.0606} & \textcolor{blue}{0.0128} &
% \textcolor{blue}{0.0822} & \textcolor{blue}{0.0203} & \textcolor{blue}{0.0621} & \textcolor{blue}{0.0589} & \textcolor{blue}{0.0185} &
% \textcolor{blue}{1.3986} & \textcolor{blue}{0.7673} & \textcolor{blue}{1.6730} & \textcolor{blue}{2.1473} & \textcolor{blue}{0.7535} \\

%  \textcolor{blue}{{TransBTS}} & \textcolor{blue}{0.0825} & \textcolor{blue}{0.0187} & \textcolor{blue}{0.0621} & \textcolor{blue}{0.0607} & \textcolor{blue}{0.0168} &
% \textcolor{blue}{0.0734} & \textcolor{blue}{0.0243} & \textcolor{blue}{0.0578} & \textcolor{blue}{0.0568} & \textcolor{blue}{0.0224} &
% \textcolor{blue}{1.2659} & \textcolor{blue}{0.7141} & \textcolor{blue}{1.4367} & \textcolor{blue}{2.1414} & \textcolor{blue}{0.6793} \\

{U-KAN}
& 0.0823 & 0.0173 & 0.0599 & 0.0600 & 0.0158
& 0.0723 & 0.0229 & 0.0560 & 0.0578 & 0.0217
& 1.0480 & 1.5814 & 1.6491 & 2.1208 & 1.5716 \\
{UKAN-EP (ECA after PFA)}
& 0.0846 & 0.0141 & 0.0638 & 0.0593 & 0.0127
& 0.0769 & 0.0119 & 0.0598 & 0.0571 & 0.0186
& 0.9484 & 1.2052 & 1.4523 & 2.0314 & 1.1800 \\
{UKAN-EP (ECA before PFA)}
& 0.0806 & 0.0176 & 0.0608 & 0.0608 & 0.0164
& 0.0723 & 0.0235 & 0.0567 & 0.0583 & 0.0225
& 1.5408 & 1.1368 & 1.2773 & 2.3427 & 1.1567 \\
{U-KAN+PFA}
& 0.0830 & 0.0165 & 0.0565 & 0.0595 & 0.0151
& 0.0746 & 0.0220 & 0.0574 & 0.0574 & 0.0207
& 1.4042 & 1.7085 & 1.5208 & 2.2508 & 1.6950 \\
{U-KAN+ECA (ECA after Conv)}
& 0.0763 & 0.0193 & 0.0652 & 0.0618 & 0.0176
& 0.0634 & 0.0251 & 0.0561 & 0.0586 & 0.0234
& 1.7735 & 1.1708 & 1.3669 & 2.4020 & 1.1494 \\
{U-KAN+ECA (ECA after skip connection)}
& 0.0830  & 0.0166 & 0.0633 & 0.0608 & 0.0157
& 0.0740  & 0.0223 & 0.0589 & 0.0580 & 0.0215
& 1.2042 & 2.2302 & 1.3176 & 2.3629 & 1.9872 \\
{U-KAN+PFA+ESA}
& 0.0738 & 0.0194 & 0.0652 & 0.0601 & 0.0127
& 0.0682 & 0.0200 & 0.0517 & 0.0496 & 0.0198
& 1.4825 & 0.8120 & 1.9725 & 2.0914 & 0.5962 \\
{U-KAN+ESA}
& 0.0810  & 0.0202 & 0.0617 & 0.0527 & 0.0186
& 0.0756 & 0.0239 & 0.0492 & 0.0653 & 0.0255
& 1.4835 & 0.7914 & 2.5936 & 2.2027 & 0.8392 \\
{U-KAN+PFA+ECA+ESA}
& 0.0827 & 0.0176 & 0.0602 & 0.0574 & 0.0165
& 0.0737 & 0.0233 & 0.0570 & 0.0561 & 0.0222
& 1.6939 & 0.8023 & 1.4497 & 2.2455 & 0.6626 \\
{U-KAN+ECA+ESA}
& 0.0860 & 0.0157 & 0.0623 & 0.0610 & 0.0141
& 0.0764 & 0.0215 & 0.0579 & 0.0583 & 0.0201
& 1.2833 & 0.8368 & 2.1394 & 2.1002 & 0.8097 \\
{U-KAN+PFA+Self-Attention}
& 0.0880 & 0.0177 & 0.0613 & 0.0601 & 0.0171
& 0.0657 & 0.0231 & 0.0557 & 0.0557 & 0.0227
& 1.5799 & 1.2955 & 1.5226 & 2.4261 & 1.2621 \\
{U-KAN+Self-Attention}
& 0.0825 & 0.0183 & 0.0624 & 0.0583 & 0.0168
& 0.0719 & 0.0236 & 0.0584 & 0.0569 & 0.0223
& 1.3969 & 0.8338 & 1.3611 & 2.2328 & 1.1331 \\
\hline
\end{tabular}
\end{sidewaystable}

\bigskip
\bigskip
\begin{sidewaystable}[h!]
\setlength{\tabcolsep}{2pt}
\caption{Average evaluation metrics on the additional test set (271 cases) from the post-challenge dataset.}\label{tab:dice_271}
\tiny
\begin{tabular}{l|ccccc|ccccc|ccccc} 
\hline
 & \multicolumn{5}{c|}{\underline{Dice}} & \multicolumn{5}{c|}{\underline{IoU}} & \multicolumn{5}{c}{\underline{HD95}} \\ 
Model & {NETC} & {SNFH} & {ET} & {RC} & {WT} 
  & {NETC} & {SNFH} & {ET} & {RC} & {WT} 
  & {NETC} & {SNFH} & {ET} & {RC} & {WT}  \\ \hline
{U-Net} &0.3540& 0.8064& 0.6917& 0.4660& 0.8180& 0.2787& 0.7280& 0.6048& 0.3862& 0.7451& 6.7213& 7.5326& 6.7170& 11.7084& 8.1482\\
{Att-Unet} &0.2761& 0.8034& 0.6927& 0.4474& 0.8146& 0.1996& 0.7270& 0.6042& 0.3668& 0.7424& 6.7932& 8.2808& 6.7674& 13.7766& 8.7867\\
{Swin UNETR} &0.3203& 0.8009& \textbf{ 0.7148}& 0.4619& 0.8159& 0.2514& 0.7229&\textbf{  0.6332}& 0.3794& 0.7430& 8.8727& 10.4492& 7.1993& 14.0716& 10.2912\\
{VT-Unet} & 0.3481 & 0.8031 & 0.6894 & 0.4855 & 0.8217 &
0.2735 & 0.7395 & 0.6057 & 0.4074 & 0.7459 &
\textbf{ 5.5367}& 7.5678 & 6.9526 & 12.5367 & \textbf{ 7.2472}\\
{TransBTS} & 0.3274 & 0.7918 & 0.6629 & 0.4349 & 0.8017 &
0.2284 & 0.7128 & 0.5892 & 0.3819 & 0.7381 &
6.6192 & 9.1837 & \textbf{ 6.7136}& 14.2911 & 9.4923 \\
{U-KAN} &0.3247& 0.8089& 0.6829& 0.4874& 0.8200& 0.2490& 0.7318& 0.5941& 0.4040& 0.7481& 5.8269& 7.4589& 7.1452& 12.0359& 8.0440\\
{UKAN-EP} &\textbf{ 0.3672}& \textbf{ 0.8177}& 0.6736&\textbf{  0.4964}&\textbf{ 0.8275}& \textbf{ 0.2821}&\textbf{ 0.7410}&0.5883&\textbf{ 0.4121}&\textbf{ 0.7557}&6.1324&\textbf{ 7.4238}&6.8256&\textbf{ 11.2958}&7.8767\\
\hline
\end{tabular}

\bigskip
\bigskip
%\begin{sidewaystable}[h!]
\setlength{\tabcolsep}{2pt}
\caption{Uncertainties (1.96 standard errors) of the average evaluation metrics on the additional test set (271 cases) from the post-challenge dataset.}\label{tab:se_271}
\tiny
\begin{tabular}{l|ccccc|ccccc|ccccc} 
\hline
 & \multicolumn{5}{c|}{\underline{Dice}} & \multicolumn{5}{c|}{\underline{IoU}} & \multicolumn{5}{c}{\underline{HD95}} \\ 
Model & {NETC} & {SNFH} & {ET} & {RC} & {WT} 
  & {NETC} & {SNFH} & {ET} & {RC} & {WT} 
  & {NETC} & {SNFH} & {ET} & {RC} & {WT}  \\ \hline
{U-Net} &0.0504& 0.0295& 0.0409& 0.0467& 0.0295& 0.0432& 0.0307& 0.0390& 0.0426& 0.0307& 1.3113& 1.8779& 1.8465& 2.1450& 1.8941\\
{Att-Unet} &0.0417& 0.0303& 0.0403& 0.0450& 0.0307& 0.0334& 0.0317& 0.0391& 0.0408& 0.0314& 1.1766& 1.9053& 1.7342& 2.1014& 1.9708\\
{Swin UNETR} &0.0470& 0.0302& 0.0401& 0.0443& 0.0297& 0.0402& 0.0315& 0.0394& 0.0403& 0.0310& 1.5264& 2.1357& 1.9160& 2.3466& 2.0520\\
{VT-Unet} & 0.0415 & 0.0329 & 0.0413 & 0.0478 & 0.0283 &
0.0425 & 0.0349 & 0.0328 & 0.0487 & 0.0382 &
1.5043 & 1.3791 & 1.4341 & 2.5487 & 1.3794 \\
{TransBTS} & 0.0435 & 0.0365 & 0.0432 & 0.0453 & 0.0278 &
0.0437 & 0.0391 & 0.0314 & 0.0485 & 0.0317 &
1.7824 & 1.3284 & 1.8467 & 2.0137 & 2.0137 \\
{U-KAN} &0.0493& 0.0297& 0.0406& 0.0458& 0.0296& 0.0415& 0.0309& 0.0387& 0.0422& 0.0308& 1.0055& 1.8680& 2.0427& 2.2215& 1.9187\\
{UKAN-EP} & 0.0501& 0.0285& 0.0417& 0.0452& 0.0285& 0.0421&0.0299&0.0397&0.0416&0.0298&1.4281&1.8871&1.9345&2.0906&1.9200\\
\hline
\end{tabular}
\end{sidewaystable}

\bigskip
\bigskip
\begin{sidewaystable}[h!]
\setlength{\tabcolsep}{2pt}
\caption{
Average evaluation metrics on the test set (135 cases)
in the ablation study. 
}\label{tab:ablation_dice}
\centering
\tiny
\begin{tabular}{l|ccccc|ccccc|ccccc}
\hline
& \multicolumn{5}{c|}{\underline{Dice}} & \multicolumn{5}{c|}{\underline{IoU}} & \multicolumn{5}{c}{\underline{HD95}} \\ 
Model & {NETC} & {SNFH} & {ET} & {RC} & {WT}
 & {NETC} & {SNFH} & {ET} & {RC} & {WT}
 & {NETC} & {SNFH} & {ET} & {RC} & {WT}  \\ 
\hline

{UKAN-EP (ECA after PFA)}
 & \textbf{0.5197} & \textbf{0.8887} & 0.6149 & 0.6924& \textbf{0.9001}
 & \textbf{0.4238} & \textbf{0.8086} & \textbf{0.5278} & 0.6156& \textbf{0.8257}
 & \textbf{2.7882} & 3.1934 & 4.8962 & \textbf{4.4771} & 3.1425\\

{UKAN-EP (ECA before PFA)}
 & 0.4721 & 0.8726 & 0.5980 & 0.6665 & 0.8853
 & 0.3765 & 0.7863 & 0.5043 & 0.5876 & 0.8051
 & 3.8994 & 3.1616 & \textbf{4.4587} & 7.2597 & 3.2326 \\

{U-KAN+PFA}
 & 0.4985 & 0.8811 & 0.6141 & 0.6820 & 0.8933
 & 0.4029 & 0.7988 & 0.5202 & 0.6055 & 0.8170
 & 4.1778 & 3.9347 & 5.2369 & 7.0456 & 3.8262 \\

{U-KAN+ECA (ECA after Conv)}& 0.3261 & 0.8544 & 0.5760 & 0.6605 & 0.8693
 & 0.2409 & 0.7639 & 0.4827 & 0.5818 & 0.7850
 & 4.6163 & 3.9866 & 5.8101 & 7.3366 & 4.0190 \\

{U-KAN+ECA (ECA after skip connection)}& 0.4684 & 0.8813 & 0.6036 & 0.6749 & 0.8919
 & 0.3781 & 0.7993 & 0.5132 & 0.5983 & 0.8154
 & 3.4778 & 5.0650& 4.5259 & 7.3148 & 4.8006 \\

{U-KAN+PFA+ESA}& 0.5023 & 0.8727 & 0.6113 & 0.6954 & 0.8866
 & 0.4018 & 0.8001 & 0.5021 & 0.6102 & 0.8097
 & 4.1034 & 3.6414 & 5.1344 & 6.8159 & 3.1957 \\

{U-KAN+ESA}& 0.4710 & 0.8710 & 0.5812 & 0.6386 & 0.8707
 & 0.3798 & 0.7801 & 0.4953 & 0.5863 & 0.8064
 & 4.3394 & 4.2285 & 5.8356 & 7.8365 & 4.3856 \\

{U-KAN+PFA+ECA+ESA}& 0.5196 & 0.8817 & \textbf{0.6193} & \textbf{0.7082} & 0.8936
 & 0.4188 & 0.8012 & 0.5238 & \textbf{0.6288} & 0.8191
 & 3.7769 & \textbf{3.1481}& 4.7362 & 6.6024 & \textbf{2.9404} \\

{U-KAN+ECA+ESA}& 0.4878 & 0.8831 & 0.6060 & 0.6784 & 0.8947
 & 0.3987 & 0.8011 & 0.5139 & 0.6027 & 0.8184
 & 3.3237 & 3.2577 & 5.6223 & 6.7090 & 3.2073 \\

{U-KAN+PFA+Self-Attention}& 0.4162 & 0.8457 & 0.5591 & 0.6070 & 0.8573
 & 0.3203 & 0.7449 & 0.4605 & 0.5169 & 0.7620
 & 4.7108 & 5.0787 & 5.9168 & 10.0001 & 5.1369 \\

{U-KAN+Self-Attention}& 0.4213 & 0.8818 & 0.5959 & 0.6902 & 0.8925
 & 0.3306 & 0.8019 & 0.5046 & 0.6096 & 0.8175
 & 4.1351 & 3.1955 & 4.5582 & 6.8603 & 3.5428 \\

\hline
\end{tabular}

\bigskip
\bigskip
\bigskip

%\begin{sidewaystable}[h!]
\setlength{\tabcolsep}{2pt}
\caption{
Average evaluation metrics
on the test set (135 cases)
for UKAN-EP models trained with
 dynamic and fixed 
loss weighting strategies.}\label{tab:twoloss}
\centering
\begin{tabular}{l|ccccc|ccccc|ccccc} 
\hline
& \multicolumn{5}{c|}{\underline{Dice}} & \multicolumn{5}{c|}{\underline{IoU}} & \multicolumn{5}{c}{\underline{HD95}} \\ 
Strategy & {NETC} & {SNFH} & {ET} & {RC} & {WT}
 & {NETC} & {SNFH} & {ET} & {RC} & {WT}
 & {NETC} & {SNFH} & {ET} & {RC} & {WT}  \\ \hline
{Dynamic  weights}& \textbf{0.5197}& \textbf{0.8887}& 0.6149& \textbf{0.6924}& \textbf{0.9001}& \textbf{0.4238}&\textbf{0.8086}&0.5278&\textbf{0.6156}&\textbf{0.8257}&\textbf{2.7882}&3.1934&4.8962&\textbf{4.4771}&\textbf{3.1425}\\
{Fixed  weights} &0.4508&0.8826&\textbf{0.6302}&0.6748&0.8944&0.3582&0.8002&\textbf{0.5340}&0.5966&0.8178&3.8816&\textbf{2.8534}&\textbf{4.4809}&7.5567&3.8299\\

\hline
\end{tabular}

\bigskip
\bigskip
\bigskip
\bigskip
\setlength{\tabcolsep}{2pt}
\caption{Uncertainties (1.96 standard errors) of the average evaluation metrics
on the test set (135 cases)
for UKAN-EP models trained with
 dynamic and fixed 
loss weighting strategies.}\label{tab:twoloss SD}
\centering
\begin{tabular}{l|ccccc|ccccc|ccccc} 
\hline
& \multicolumn{5}{c|}{\underline{Dice}} & \multicolumn{5}{c|}{\underline{IoU}} & \multicolumn{5}{c}{\underline{HD95}} \\ 
Strategy & {NETC} & {SNFH} & {ET} & {RC} & {WT}
 & {NETC} & {SNFH} & {ET} & {RC} & {WT}
 & {NETC} & {SNFH} & {ET} & {RC} & {WT}  \\ \hline
{Dynamic  weights}
& 0.0846 & 0.0141 & 0.0638 & 0.0593 & 0.0127
& 0.0769 & 0.0119 & 0.0598 & 0.0571 & 0.0186
& 0.9484 & 1.2052 & 1.4523 & 2.0314 & 1.1800 \\
{Fixed  weights}& 0.0751 & 0.0182 & 0.0609 & 0.0571 & 0.0198
& 0.0835 & 0.0273 & 0.0473 & 0.0581 & 0.0209
& 1.2794 & 1.9437 & 1.2764 & 2.5673 & 1.1089 \\

\hline
\end{tabular}
\end{sidewaystable}

\subsection{Ablation Study}\label{sec:result_ablation}

We conduct a comprehensive ablation study to evaluate the contributions of individual components in the proposed UKAN-EP (ECA after PFA), including the effects of ECA and PFA modules, ESA and self-attention alternatives, ViT block integration into U-KAN, and the design of the loss function.%Quantitative results are provided in Table~\ref{tab:ablation_dice}, and model stability is assessed in Table~\ref{tab:SD}.

\subsubsection{Roles of ECA, PFA, and Other Attention Mechanisms}

We assess the contributions of the ECA and PFA modules, along with alternative attention mechanisms, using a series of ablation experiments. 
Table~\ref{tab:ablation_dice} shows the 
average evaluation metrics,
with uncertainties
given in Table~\ref{tab:SD}.

\medskip
\noindent\textbf{Effect of ECA.}  
To evaluate the impact of ECA placement, we compare the default configuration, where ECA follows PFA, with a variant where ECA is applied before the PFA module. This corresponds to the UKAN-EP (ECA after PFA) vs. UKAN-EP (ECA before PFA) comparison in Table~\ref{tab:ablation_dice}. This change results in a noticeable performance decline across all average metrics, except the average HD95 for ET and SNFH. The results demonstrate the importance of applying ECA after PFA to achieve better channel recalibration. Retaining the PFA module while removing the ECA (i.e., U-KAN+PFA) noticeably reduces the average Dice score from 0.5197 to 0.4985 for NETC and from 0.6924 to 0.6820 for RC, compared to UKAN-EP (ECA after PFA). The average IoU scores drop similarly, and the average HD95 increases substantially for NETC from 2.7882 to 4.1778 and for RC from 4.4771 to 7.0456. 

\medskip
\noindent\textbf{Effect of PFA.}  
To gauge the effectiveness of the  PFA module,
we remove it and retain only the ECA module, placing ECA either after each encoder convolutional layer 
(U-KAN+ECA (ECA after Conv)) or after each skip connection (U-KAN+ECA (ECA after skip connection)). For the U-KAN+ECA (ECA after Conv) configuration, the average Dice scores drop sharply to 0.3261 for NETC, 0.5760 for ET, and 0.6605 for RC. The performance improves  when ECA is placed after the skip connection, but remains inferior to UKAN-EP (ECA after PFA). This highlights the critical role of PFA in enabling spatial context fusion and improving regional precision.

\medskip
\noindent\textbf{ECA vs. ESA and Self-Attention.}  
To evaluate the contribution of other types of attention mechanisms, we consider multiple configurations incorporating Efficient Spatial Attention (ESA)~\citep{9460078} and self-attention~\citep{vaswani2017attention}. ESA follows the strategy of ECA, replacing channel-wise weighting with spatial-wise weighting. First, the U-KAN+PFA+ESA variant, which replaces ECA with ESA in UKAN-EP (ECA after PFA), yields slightly weaker  performance in all metrics except the average Dice for RC. 
When both ESA and ECA are utilized (i.e., U-KAN+PFA+ECA+ESA), we observe slight improvements, with the highest
average Dice scores for RC (0.7082) and ET (0.6193), highest average IoU for RC (0.6288), and lowest average HD95 for SNFH (3.1481) and WT (2.9404), as shown in Table~\ref{tab:ablation_dice}. 
The proposed UKAN-EP (ECA after PFA) achieves comparable scores
in these metrics,
but
 outperforms U-KAN+PFA+ECA+ESA on all other average metrics except  a slightly higher average HD95 for ET (4.8962 vs. 4.7362),
and notably achieves
 much lower average
 HD95 for NETC (2.7882 vs. 3.7769)
 and RC (4.4771 vs. 6.6024). 
Removing the PFA module 
and placing the attention mechanisms after each encoder convolutional layer 
(i.e., U-KAN+ESA and U-KAN+ESA+ECA) 
still leads to inferior performance 
compared to the proposed UKAN-EP (ECA after PFA) across all metrics.
Replacing them with the self-attention mechanism 
 (U-KAN+Self-Attention) also underperforms relative to our model.  Moreover, adding the PFA module before  the self-attention mechanism (U-KAN+PFA+Self-Attention) results in further performance degradation across all metrics. This suggests that while ESA and self-attention  can provide complementary benefits, the ECA and PFA modules have the most significant impact on segmentation accuracy, underscoring the effectiveness of channel-wise recalibration following multi-scale feature aggregation.

\subsubsection{Integration of ViT into U-KAN}
\begin{figure}[t!]
    \centering
    % First row
\subfigure[ViT used as a full replacement for the CNN encoder\label{fig:111}]{
        \includegraphics[width=0.485\textwidth]{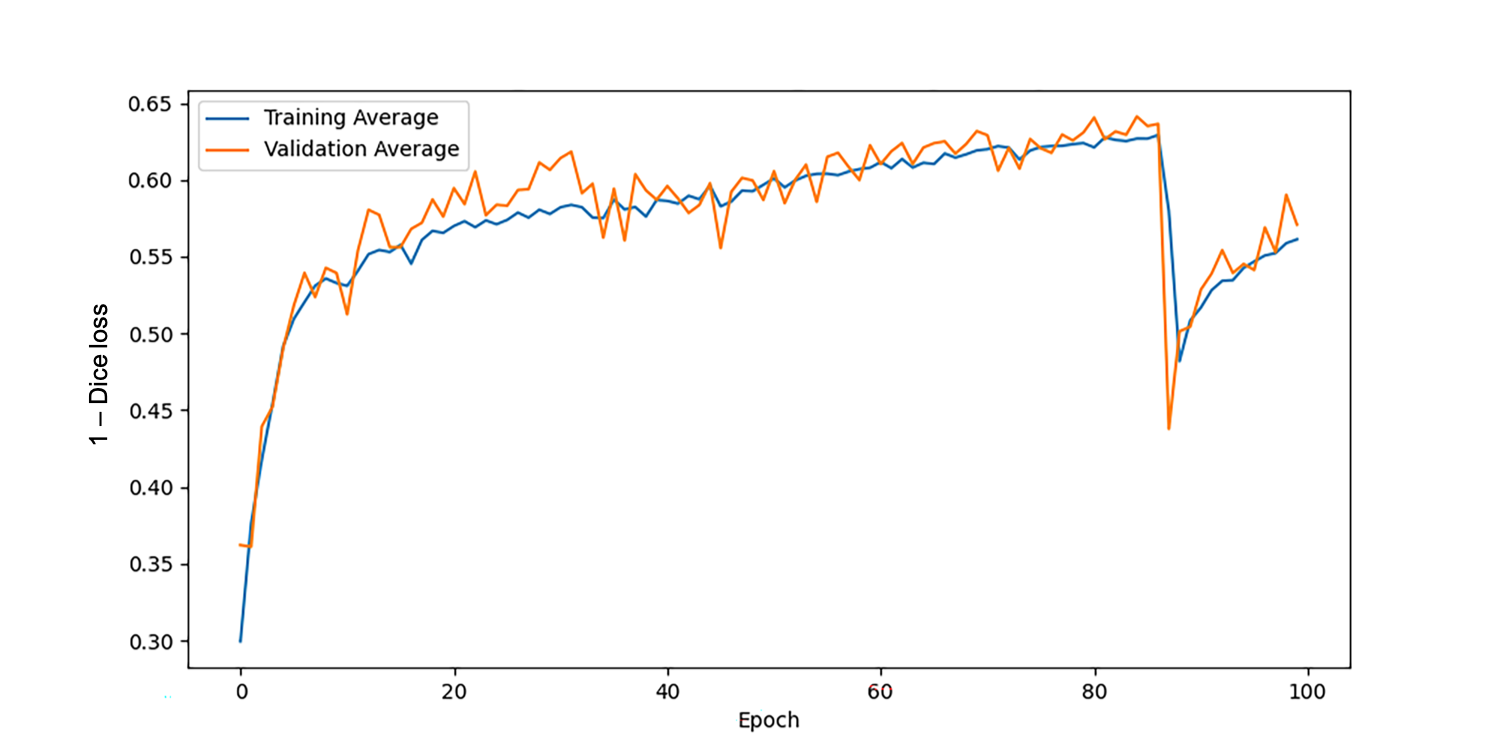}}
    \hfill
        \subfigure[ViT inserted between the CNN encoder and the Tok-KAN bottleneck \label{fig:333}]{
        \includegraphics[width=0.485\textwidth]{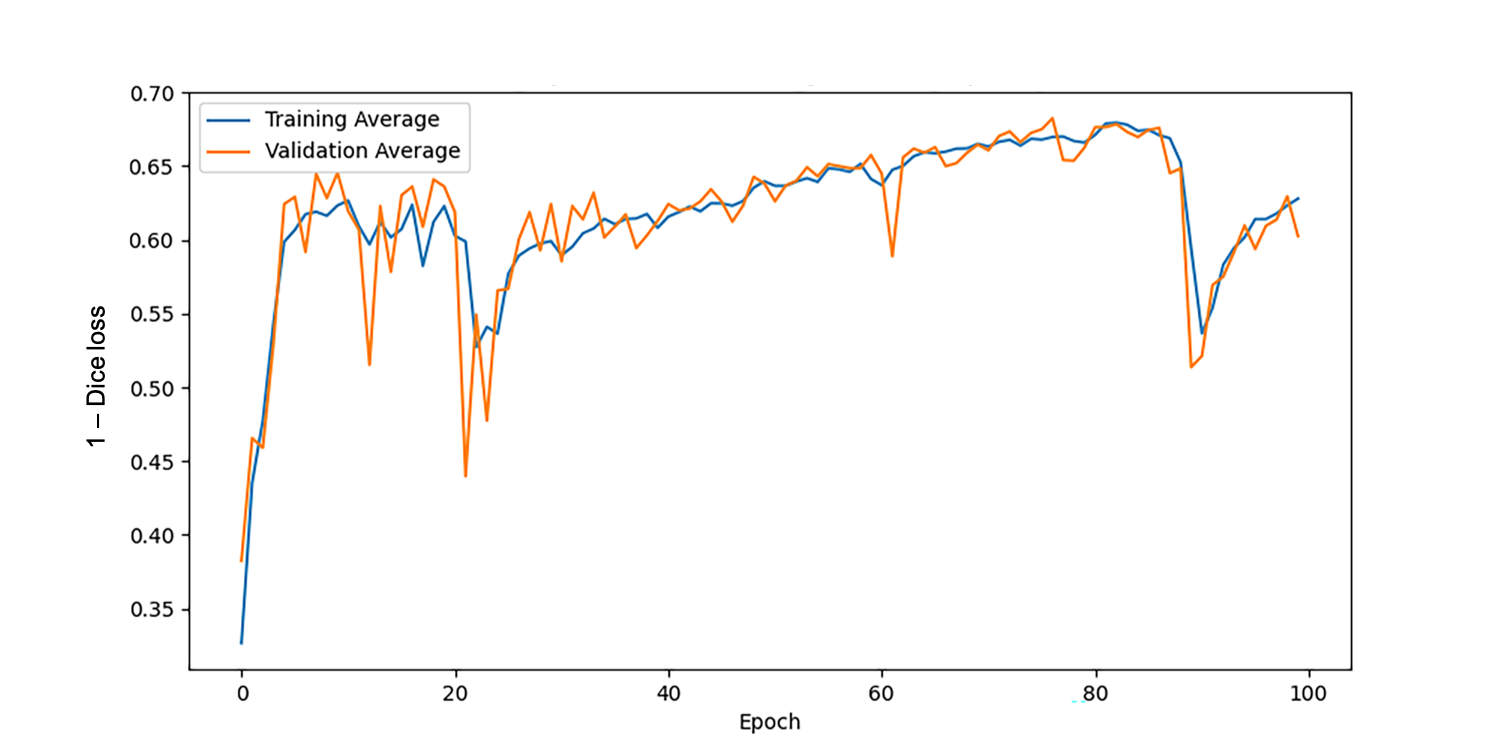}}

    % Second row
    \vspace{1em}
\subfigure[ViT inserted between the two lowest-level Tok-KAN blocks\label{fig:222}]{
        \includegraphics[width=0.485\textwidth]{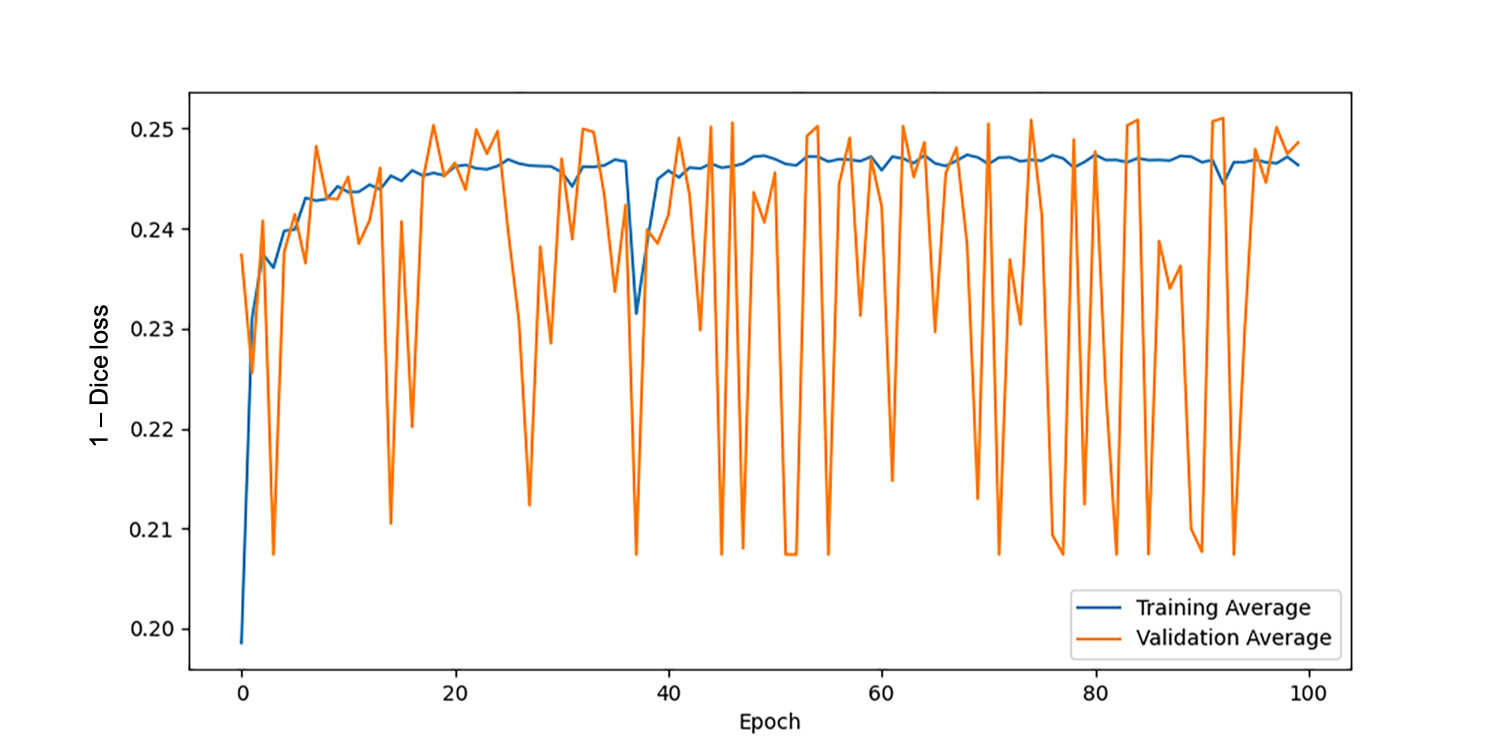}}
    \hfill
\subfigure[U-KAN without ViT\label{fig:ukanloss}]{
        \includegraphics[width=0.485\textwidth]{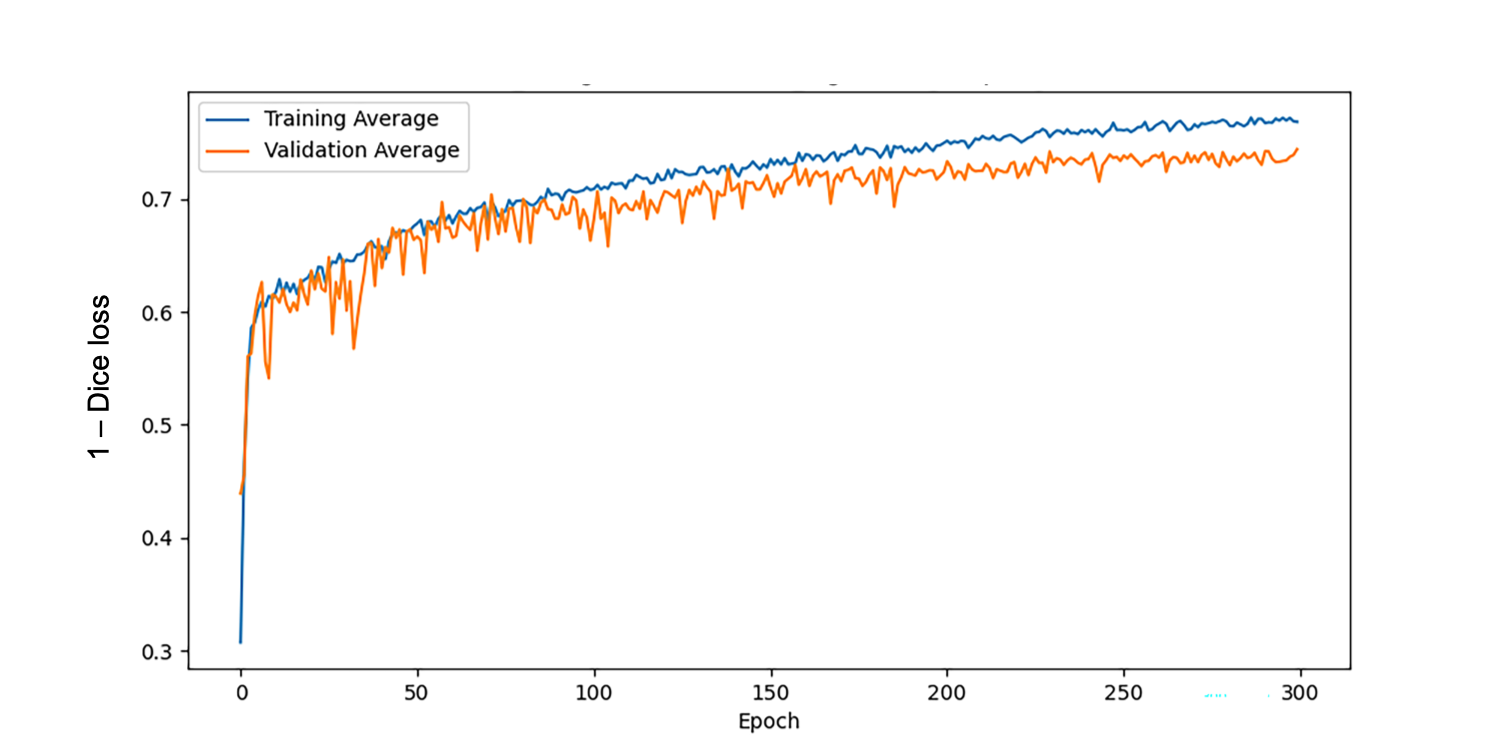}}
    % Main title
    \caption{Comparison of     
overall soft Dice scores (i.e., 1 $-$ Dice loss; see \eqref{individual dice loss})  averaged separately over the training and validation sets during U-KAN training with different ViT configurations.}
    \label{fig:kan_transformer_comparison}
\end{figure}

While the previous subsection considered standalone self-attention,  we now evaluate the integration of full transformer components.
Recent studies have demonstrated the advantages of incorporating KAN layers into transformers for vision tasks,
either by replacing only the MLP layers~\citep{yang2024kolmogorov}
or by substituting both the 
MLP layers and the QKV mapping matrices~\citep{wu2024transukan}.
To  examine whether similar benefits apply to the U-KAN architecture for 3D image segmentation, we consider integrating a Vision Transformer (ViT) block \citep{dosovitskiy2020vit} at different locations within U-KAN. 
Unlike standalone self-attention, a ViT block consists of four transformer encoder layers, each comprising a multi-head self-attention mechanism followed by an MLP. 
We explore three integration strategies:
(i)
replacing the entire CNN encoder with a ViT block
 to enable global context modeling during feature extraction; (ii) 
inserting a ViT block between the CNN encoder and the Tok-KAN bottleneck to assess its effect  on high-level semantic representation; and
(iii)
 placing a ViT block between the two lowest-level Tok-KAN blocks  to capture long-range dependencies before decoding, with the ViT output fused with the original feature map to combine global context and local detail.
 These configurations allow a comprehensive 
 evaluation of the ViT block’s impact at different locations within U-KAN. As shown in Figure~\ref{fig:111}, using a ViT-based  encoder  reduces the
average overall soft Dice score compared to the original CNN-based encoder in U-KAN (Figure~\ref{fig:ukanloss}).
 Inserting the ViT block between
 the CNN encoder and the
Tok-KAN bottleneck (Figure~\ref{fig:333}) yields no significant performance gains and sometimes  causes sharp drops during training. 
When the ViT block is  placed between the two lowest-level Tok-KAN blocks (Figure~\ref{fig:222}), the average  overall soft Dice  on the validation set 
remains  very low (below 0.26) and
shows large fluctuations throughout training. 
Furthermore,
these ViT-based variants introduce substantial computational overhead. In contrast, as illustrated in Figure~\ref{fig:ukanloss}, the 3D U-KAN without ViT consistently achieves superior and more stable segmentation performance, highlighting the limited benefit of
directly integrating the ViT block
into the U-KAN architecture.

\subsubsection{Loss Function Design}\label{sec:abl_loss}
We evaluate the proposed dynamic loss weighting strategy (Section~\ref{sec: loss function}) against a fixed weighting strategy defined as
 $\mathcal{L}_{\text{total}} = \frac{1}{B}\sum_{i=1}^B\{0.5 \mathcal{L}_{\text{CE}}^{(i)} + 0.5\mathcal{L}_{\text{Dice}}^{(i)}\}$. 
 Tables~\ref{tab:twoloss} and~\ref{tab:twoloss SD} present the segmentation performance and associated uncertainties for both strategies. 
The dynamic strategy  achieves overall superior results, notably improving NETC segmentation by 15.28\% in average Dice, 18.31\% in average IoU, and reducing average HD95 by 28.17\%. 
Uncertainties are comparable between the two strategies,
except for SNFH, where fixed weighting results in  
129\%
and 
61.3\% higher uncertainty in average IoU and average HD95, respectively.
These improvements underscore the effectiveness of dynamic weighting in enhancing both overlap accuracy and boundary precision across tumor subregions.

\iffalse
\begin{figure}[H]
    \centering
    \includegraphics[width=\textwidth]{twoloss.png}
    \caption{Loss values for ce loss and dice loss}
    \label{fig:twoloss}
\end{figure}
\fi

\subsection{Computational Efficiency}\label{sec:result_comp}

Table~\ref{tab:param_time} reports the computational complexity of each model in terms of Giga Floating Point Operations (GFLOPs) and the number of trainable parameters (in millions). U-KAN demonstrates lower computational overhead than U-Net, achieving a 9\% reduction in GFLOPs (107.71 vs. 118.96) and a 40\% reduction in parameters (10.61M vs. 17.56M), while maintaining comparable segmentation performance. Extending U-KAN with the proposed PFA and ECA modules moderately increases the computational cost; UKAN-EP requires 223.57 GFLOPs and 11.30M parameters, but presents significant  gains in segmentation performance as reported in Section~\ref{sec:result_seg_perf}. 

Compared to Att-Unet and Swin UNETR, UKAN-EP is significantly more efficient. It reduces GFLOPs by 68\% (223.57 vs. 708.44) with a moderate increase in parameter count (11.30M vs. 6.44M) relative to Att-Unet, and reduces GFLOPs by 88\% (223.57 vs. 1846.21) and parameter count by 82\% (11.30M vs. 62.36M) when compared to Swin UNETR. Compared to U-KAN+Self-Attention, UKAN-EP incurs an 11\% increase in GFLOPs (223.57 vs. 201.10) and a 1.4\% increase in parameters (11.30M vs. 11.14M), yet delivers superior segmentation accuracy and boundary precision. 
%{\color{black}
Compared to VT-UNet, UKAN-EP reduces GFLOPs by 82\% (223.57 vs. 1259.46) and parameter count by 63\% (11.30M vs. 30.80M). Compared to TransBTS, UKAN-EP reduces GFLOPs by 55\% (223.57 vs. 498.52) and parameter count by 91\% (11.30M vs. 126.8M). These results further demonstrate the favorable trade-off offered by UKAN-EP between segmentation performance and computational efficiency.
%}

\begin{table}[h!]
\centering
\caption{GFLOPs and number of total parameters for each model.}
{\begin{tabular}{lcc}
\hline
{Model} & {GFLOPs} & {Params (M)} \\ 
\hline
{U-Net}                    & 118.96  & 17.56  \\
{Att-Unet}                 & 708.44  & 6.44   \\
{Swin UNETR}               & 1846.21 & 62.36  \\
 {VT-Unet}                 & 1259.46&30.80\\
{TransBTS}                 &498.52 &126.80\\
 %  \textcolor{blue}{{VT-Unet}}& \textcolor{blue}{1259.46}&\textcolor{blue}{30.80}\\
 % \textcolor{blue}{{TransBTS}}&\textcolor{blue}{498.52}&\textcolor{blue}{126.80}\\
{U-KAN}                    & 107.71  & 10.61  \\
{UKAN-EP (ECA after PFA)}                  & 223.57  & 11.30  \\ 
{UKAN-EP (ECA before PFA)} & 223.67  & 11.30  \\
{U-KAN+PFA}                 & 223.54  & 11.30  \\
{U-KAN+ECA (ECA after Conv)}& 194.56   & 10.61  \\
{U-KAN+ECA (ECA after skip connection)}& 200.12& 11.14\\
{U-KAN+PFA+ESA}& 223.52& 11.30\\
{U-KAN+ESA}& 200.07& 11.30\\
{U-KAN+PFA+ECA+ESA}& 223.55& 11.30\\
{U-KAN+ECA+ESA}& 200.10& 11.30\\
{U-KAN+PFA+Self-Attention}& 224.55& 11.30\\
{U-KAN+Self-Attention}& 201.10& 11.14\\
\hline
\end{tabular}}
\label{tab:param_time}
\end{table}

%{\color{black}
\section{Discussion}\label{sec:discussion}

\subsection{Advances over Existing Models}

Compared with conventional U-Net, which captures multi-scale features through its encoder-decoder structure, our proposed UKAN-EP incorporates Tok-KAN blocks together with PFA and ECA modules to enhance both interpretability and multi-scale feature fusion. While Att-Unet augments the U-Net design with attention gates to better highlight relevant regions, UKAN-EP further strengthens feature recalibration via lightweight and efficient channel attention. Swin UNETR leverages a Swin Transformer encoder to capture hierarchical dependencies, but it requires substantially larger computational complexity (both parameters and GFLOPs). In contrast, UKAN-EP achieves competitive or superior segmentation performance with lower computational cost. Recently proposed transformer-based models such as TransBTS and VT-UNet also demonstrate strong performance, but incur significantly higher GFLOPs and parameter counts. UKAN-EP balances efficiency and accuracy, making it a practical alternative for 3D tumor segmentation.

\subsection{Limitations and Future Work}

Despite these advances, our study has several limitations. First, the experiments were primarily conducted on the BraTS post-treatment dataset, and thus the generalizability of the proposed model to other clinical datasets requires further validation. Second, although UKAN-EP achieves strong overall performance, its ET  segmentation accuracy is occasionally slightly lower than that of Swin UNETR. This may be due to the hierarchical attention mechanism in Swin UNETR, which more effectively captures long-range contextual dependencies that are particularly important for ET regions. Addressing this limitation will be an important direction for future work, for example by incorporating stronger global context modeling into UKAN-EP. 

In future work, we aim to extend UKAN-EP to larger, more heterogeneous datasets to evaluate its stability across different scanners, protocols, and patient populations. Another promising direction is to adapt the framework to other 3D tumor segmentation tasks beyond gliomas, enabling broader clinical applicability. Finally, integrating complementary modalities such as genomics or digital pathology with MRI could unlock richer multi-modal representations, potentially enhancing downstream tasks such as tumor classification, treatment response prediction, and survival analysis.
%}

\section{Conclusion}\label{sec: conclusion}
This study presents UKAN-EP, a novel 3D extension of the original 2D U-KAN model, which integrates ECA and PFA modules for multi-modal MRI brain tumor segmentation.
The proposed UKAN-EP  is evaluated on the 2024 BraTS-GLI dataset and demonstrates strong segmentation performance with high computational efficiency. Compared to self-attention-based models such as Att-Unet and Swin UNETR, UKAN-EP achieves better segmentation performance while requiring only a fraction of the computational cost. Although transformer-based models are effective at modeling long-range dependencies in large-scale vision tasks, we find they perform less reliably in small-sample 3D medical image segmentation. In contrast, UKAN-EP consistently delivers robust performance using only basic data augmentation, highlighting the impact of PFA and ECA modules that enhance skip connections through multi-scale spatial aggregation and channel-wise recalibration.

\section*{List of abbreviations}\label{sec:List_of_abbreviations}

\begin{tabular}{@{}ll@{}}
\toprule
\textbf{Abbreviation} & \textbf{Full Term} \\
\midrule
BraTS  & Brain Tumor Segmentation Challenge\\
MLP    & Multilayer Perceptrons          \\
CNN    & Convolutional Neural Network    \\
MRI    & Magnetic Resonance Imaging      \\
T1     & T1-weighted                     \\
T1Gd   & Contrast-enhanced T1-weighted   \\
T2     & T2-weighted                     \\
FLAIR  & T2-weighted fluid-attenuated inversion recovery \\
ET     & Enhancing tissue                \\
NETC   & Non-enhancing tumor core        \\
RC     & Resection cavity                \\
SNFH   & Surrounding non-enhancing FLAIR hyperintensity \\
WT     & Whole tumor (ET+SNFH+NETC)      \\
KAN    & Kolmogorov-Arnold Network       \\
ECA    & Efficient Channel Attention     \\
PFA    & Pyramid Feature Aggregation     \\
ESA    & Efficient Spatial Attention     \\
ViT    & Vision Transformer              \\
Tok-KAN& Tokenized KAN                   \\
HD95   & 95th percentile Hausdorff Distance \\
IoU    & Intersection over Union         \\
GFLOPs & Giga Floating
Point Operations \\
\bottomrule
\end{tabular}

\section*{Acknowledgement}\label{sec:acknowledgement}
 This work was supported in part through the NYU IT High Performance Computing resources, services, and staff expertise.

\section*{Data Availability Statement}
The data used in this publication were obtained as part of the Challenge project through Synapse ID (syn53708249).
The data are available at
\url{https://www.synapse.org/Synapse:syn53708249/wiki/627759}.

\section*{Disclosure Statement}
The authors report there are no competing interests to declare.

\section*{Funding}
This work was partially supported by funding to Dr. Hai Shu from the New York University (NYU) GPH Research Support Grant, the NYU GPH Goddard Award, and the National Institutes of Health (NIH) grant RF1AG098697. The content is solely the responsibility of the authors and does not necessarily represent the official views of NYU or NIH. 

\section*{Ethics, Consent to Participate, and Consent to Publish Declarations}
Not applicable.


\bibliography{sn-bibliography}% common bib file

\end{document}